\documentclass[10pt,conference]{IEEEtran}
\usepackage[T1]{fontenc}
\usepackage[utf8]{inputenc}
\usepackage{booktabs,multirow,graphicx,amsmath,array,xcolor,siunitx,hyperref,microtype}
\usepackage{tikz}
\usetikzlibrary{arrows.meta,positioning,shapes.geometric}
\hypersetup{hidelinks}
\title{NPU Accelerator: Quantized Real-Time Vehicle Detection on PYNQ-Z1 Using FINN}
\author{\IEEEauthorblockN{Daniel Gutiérrez, Antonio Cuesta, Jorge Fe, Bruno Gutierrez}
\IEEEauthorblockA{\textit{Intigia}}
\and
\IEEEauthorblockN{Rashed Al Koutayni}
\IEEEauthorblockA{\textit{DFKI}}}
\begin{document}
\maketitle

\begin{abstract}
This paper presents the design, optimization, implementation, and on-board validation of a neural processing unit (NPU) accelerator for real-time vehicle detection on the resource-constrained Xilinx Zynq XC7Z020 device of the PYNQ-Z1 board. The work follows a hardware/software co-design methodology that combines quantization-aware training (QAT), lightweight YOLO-derived detectors, Brevitas/QONNX model export, FINN dataflow compilation, Vivado implementation, and physical benchmarking on the target board. Four simultaneous engineering requirements define successful deployment: throughput above 30 frames/s (FPS), energy efficiency above 7 FPS/W, programmable-logic (PL) hardware latency below 50 ms, and Pascal VOC detection accuracy above 0.55 mAP@0.5. The design space includes LP-YOLO and LP-YOLO Slim variants, a custom YOLOv3-tiny reference, 4-bit and mixed low-bit quantization, 320$\times$320 and 256$\times$256 inputs, manual and automatic FIFO sizing, and programmable-logic clocks from 100 to 200 MHz. The final LP-YOLO Slim configuration uses a 256$\times$256 input, w2a4 quantization, and a 142.86 MHz PL clock. With batch 100 it reaches 35.66 FPS at 2.91 W, corresponding to 12.25 FPS/W, while measured PL latency is 45.11 ms and VOC mAP@0.5 is 0.594. This is the only evaluated configuration for which the supplied measurements satisfy all four requirements simultaneously. The results show that low-bit QAT, architectural slimming, FINN folding and FIFO optimization, and moderate clock scaling can jointly provide a practical real-time detector on a small Zynq FPGA.
\end{abstract}


\section{Introduction}
Real-time object detection at the edge is constrained by four competing quantities: detection accuracy, throughput, inference latency, and energy consumption. High-capacity convolutional neural networks can provide strong accuracy but are difficult to deploy on small embedded FPGAs because their weights, intermediate feature maps, and arithmetic parallelism compete for limited BRAM, LUTs, DSPs, routing, and memory bandwidth. Conversely, aggressively reducing a detector may make the accelerator fast enough while degrading recall and mean average precision (mAP). The central problem is therefore not to maximize a single metric, but to identify an operating point that satisfies all system requirements at once.

The target platform in this work is the PYNQ-Z1, based on the Xilinx Zynq XC7Z020. The device combines a dual-core ARM processing system (PS) with programmable logic (PL), making it suitable for hardware/software co-design. The convolutional part of the detector is mapped to a streaming accelerator in the PL, while software-side operations such as output decoding and non-maximum suppression (NMS) remain on the ARM processor. This division concentrates FPGA resources on the regular, highly parallel convolutional workload while retaining flexibility for control-heavy post-processing.

The project initially investigated YOLOv3-tiny-derived detectors. The full model has approximately 8.66 M parameters and 1.61 GMAC per 320$\times$320 inference. At 4-bit weight precision, its parameters require about 34.7 Mbit, approximately seven times the 4.9 Mbit on-chip BRAM capacity reported for the XC7Z020 in the project analysis. A fully on-chip FINN streaming implementation is therefore impractical for that model. In contrast, LP-YOLO contains approximately 0.35 M parameters and 0.091 GMAC, with about 1.4 Mbit of 4-bit weights. This reduction makes it a feasible basis for FINN-based streaming acceleration and motivates the model-slimming direction pursued here.

The work makes three main contributions. First, it presents an end-to-end deployment methodology from quantized training to physical FPGA measurements, explicitly distinguishing FINN analytical estimates from on-board results. Second, it studies several implementation levers---model size, input resolution, quantization, PE/SIMD folding, FIFO depth, and clock frequency---and relates them to measured throughput, latency, power, and accuracy. Third, it identifies a final LP-YOLO Slim implementation at 142.86 MHz that closes all four project KPIs simultaneously, whereas the neighboring 125 MHz implementation misses the PL-latency requirement by only 1.49 ms.

\section{System Objectives and Target Platform}
\subsection{Four simultaneous KPIs}
A design is considered successful only if the same implementation satisfies the following four conditions:
\begin{equation}
\mathrm{FPS}>30,\qquad \frac{\mathrm{FPS}}{P}>7\;\mathrm{FPS/W},
\end{equation}
\begin{equation}
L_{\mathrm{HW-PL}}<50\;\mathrm{ms},\qquad \mathrm{mAP@0.5}_{VOC}>0.55.
\end{equation}
These requirements deliberately combine algorithmic and hardware criteria. A model with high mAP but insufficient FPS is not a real-time solution; a fast model with poor mAP is not an acceptable detector; and a design with adequate throughput may still violate single-image latency or power-efficiency constraints.

\subsection{PYNQ-Z1 hardware/software partition}
The PYNQ-Z1 provides the Zynq PS for software orchestration and the XC7Z020 PL for acceleration. In the FINN deployment used in this project, quantized convolutional layers form the FPGA streaming datapath. Input/output transfers, driver control, detection decoding, and NMS execute on the PS. This distinction is important when interpreting latency: \emph{HW-PL latency} isolates the accelerator, whereas \emph{end-to-end latency} also includes software and data-transfer overhead.

The final evaluation uses real board measurements. This is essential because FINN cycle estimates do not include every system overhead, and post-route timing, DMA behavior, and PS-side processing can create a measurable gap between analytical predictions and deployed performance.

\section{Model and Hardware Design Space}
\subsection{Memory and compute feasibility}
Before physical implementation, candidate networks were screened against the XC7Z020 resource envelope. This step is necessary because a high software-side mAP does not imply that the corresponding model can be converted into a practical FINN streaming accelerator. The full YOLOv3-tiny reference requires approximately 8.66 M parameters and 1.61 GMAC per 320$\times$320 inference. Under the 4-bit weight representation used in the project analysis, the parameter storage is approximately 34.7 Mbit (4.34 MB), while the available on-chip BRAM capacity is approximately 4.9 Mbit. The mismatch is roughly sevenfold. One 512-to-1024 convolution alone accounts for about 18.9 Mbit of weights, making it impossible to solve the problem only through small FIFO or folding adjustments.

LP-YOLO changes this feasibility picture. At 320$\times$320 it is reported at approximately 0.35 M parameters and 0.091 GMAC, making it about 25 times smaller in parameters and 18 times smaller in MAC count than the full YOLOv3-tiny reference. At 4-bit precision its weights occupy approximately 1.4 Mbit, comfortably below the on-chip BRAM budget. This is the architectural reason that LP-YOLO can be mapped as a FINN streaming accelerator with on-chip weights while the full reference requires a different memory strategy such as DDR streaming.

The project also evaluated intermediate width/bit-width operating points. The resulting design-space study shows that fitting the weight-memory budget and achieving the desired accuracy are distinct constraints. Some wider configurations improve mAP but exceed the memory budget; narrower configurations fit easily but lose accuracy. The hardware implementation stage therefore focuses on candidates that are both computationally plausible and sufficiently accurate to justify physical synthesis.

\subsection{From analytical estimates to physical measurements}
FINN estimates are useful for pruning the design space, but they are not treated as final performance results. The project observed this directly for LP-YOLO: a 4/4-bit folding associated with a 38 FPS FINN clock-cycle estimate measured only about 22 FPS in the corresponding 100 MHz on-board batch experiment. The analytical estimate counts accelerator cycles and is valuable for comparing foldings, but the deployed system also contains DMA transfers, stream startup/drain effects, PS-side operations, and implementation-dependent overheads. For this reason, the methodology distinguishes three levels of evidence: model-side accuracy and compute estimates, FINN cycle/resource estimates, and finally routed/on-board measurements.

This distinction also affects resource interpretation. FINN's analytical estimates guide PE/SIMD selection, but the post-route Vivado report is the authority for whether the design actually fits and meets timing. Routing congestion and timing closure can prevent a theoretically attractive parallelization from becoming a valid bitstream. The workflow therefore iterates between model selection and hardware mapping rather than treating FPGA compilation as a final mechanical conversion step.

\subsection{Optimization variables}
The final exploration can be viewed as a sequence of four coupled variables. The first is \emph{network capacity}, which controls the fundamental accuracy/compute trade-off. The second is \emph{numerical precision}, which controls weight storage and the hardware cost of arithmetic. The third is \emph{dataflow parallelism and buffering}, represented by PE/SIMD folding and FIFO depth. The fourth is \emph{clock frequency}, which scales the rate of an already mapped pipeline but also affects power and timing closure. The experiments reported in this paper intentionally exercise all four variables rather than attributing the final result to clock frequency alone.

\section{Methodology}
\subsection{End-to-end workflow}
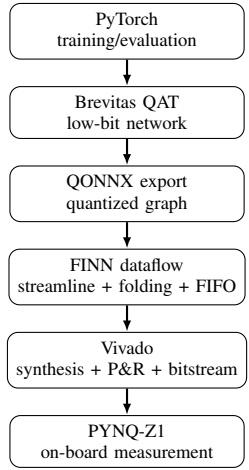
\begin{figure}[t]
\centering
\begin{tikzpicture}[node distance=3.5mm, every node/.style={font=\scriptsize}, box/.style={draw,rounded corners,align=center,minimum width=31mm,minimum height=6mm}, arr/.style={-{Latex[length=1.5mm]},thick}]
\node[box] (pt) {PyTorch\\training/evaluation};
\node[box,below=of pt] (br) {Brevitas QAT\\low-bit network};
\node[box,below=of br] (qo) {QONNX export\\quantized graph};
\node[box,below=of qo] (fi) {FINN dataflow\\streamline + folding + FIFO};
\node[box,below=of fi] (vi) {Vivado\\synthesis + P\&R + bitstream};
\node[box,below=of vi] (py) {PYNQ-Z1\\on-board measurement};
\draw[arr] (pt)--(br); \draw[arr] (br)--(qo); \draw[arr] (qo)--(fi); \draw[arr] (fi)--(vi); \draw[arr] (vi)--(py);
\end{tikzpicture}
\caption{Hardware/software co-design and deployment flow used in the project.}
\label{fig:flow}
\end{figure}

Figure~\ref{fig:flow} summarizes the workflow. Models are trained and evaluated in PyTorch, quantization is introduced with Brevitas, and the trained network is exported as a quantized ONNX/QONNX graph. FINN then transforms the graph into a streaming dataflow architecture, selects or applies folding parameters, sizes FIFOs, generates HLS/RTL components, and produces a design for Vivado. Vivado performs synthesis, implementation, place-and-route, timing analysis, and bitstream generation. The resulting accelerator is loaded on the PYNQ-Z1 and controlled from Python for physical benchmarking.

\subsection{Quantization-aware training}
Low precision is a prerequisite for fitting useful convolutional parallelism into the XC7Z020. Brevitas is used to represent quantized weights and activations during training so that the optimization process adapts to quantization error. The project evidence on the full-width YOLOv3-tiny reference shows why QAT was selected over naive post-training quantization (PTQ): at 4-bit precision, QAT recovers to within approximately 0.006 mAP@0.5 of the floating-point baseline, whereas PTQ loses approximately 0.17 mAP on the same reference setup. Thus, QAT is not merely a toolchain convenience; it is an accuracy-preservation mechanism required by the low-bit hardware budget.

Two low-precision configurations are relevant to the implementations reported here. LP-YOLO uses uniform w4a4 arithmetic, i.e., 4-bit weights and 4-bit activations. LP-YOLO Slim uses w2a4, reducing internal weights to 2 bits, using 4-bit activations, and retaining an 8-bit output interface. Reducing weight precision lowers storage and arithmetic complexity and can free FPGA resources for parallel compute and buffering. The exact operation count of a fixed network does not change with bit width, but the hardware cost per operation and the achievable parallelism do.

\subsection{Architecture and input-resolution reduction}
Quantization alone does not solve the model-capacity problem. The project therefore evaluates architectural reduction and input-size reduction as complementary optimization axes. LP-YOLO is a lightweight, single-detection-scale architecture designed to fit the streaming resource budget. LP-YOLO Slim retains the same lightweight direction but reduces the input from 320$\times$320 to 256$\times$256. Since convolutional cost scales approximately with spatial area for fixed channels and kernels, this change reduces the number and size of feature-map operations throughout the backbone.

The custom YOLOv3-tiny implementation is retained as an accuracy-oriented comparison. It demonstrates the opposite side of the trade-off: greater model capacity improves mAP but produces substantially lower throughput and higher latency on the same class of FPGA.

\subsection{FINN dataflow compilation}
FINN converts the quantized graph into a streaming accelerator in which layers exchange data through streams and FIFOs. The main compilation stages used in the project include QONNX-to-FINN conversion, graph tidy-up, streamlining, conversion to hardware layers, creation of the dataflow partition, target-FPS parallelization, specialization, HLS/RTL code generation, FIFO-depth configuration, stitched-IP creation, Vivado implementation, and PYNQ deployment generation.

The dataflow architecture exposes an explicit area/performance trade-off. Matrix-vector and convolutional units can be replicated through processing elements (PEs) and widened through SIMD lanes. Higher PE/SIMD values process more channels or operations in parallel and reduce folding, increasing potential throughput. However, the cost is higher LUT utilization, more demanding routing, and potentially larger buffering. Consequently, maximum parallelism is not always routable or timing-clean on the XC7Z020.

\subsection{FIFO optimization}
Inter-layer FIFOs decouple producer and consumer layers. Their depth influences whether short-term rate mismatch or pipeline startup behavior causes stalls. Two strategies are evaluated. In the manual configuration, FINN is explicitly given an inter-layer FIFO depth of 6. In the automatic configuration, FINN determines FIFO depths from its optimization procedure. The measured results show that FIFO configuration can change sustained throughput even at the same nominal clock and model precision, confirming that buffering is part of performance tuning rather than only a storage detail.

\subsection{Clock-frequency scaling}
Clock scaling is used after obtaining a valid accelerator structure. The LP-YOLO w4a4 design is compared at 100 and 200 MHz with manual FIFO depth 6, while LP-YOLO Slim is compared at 125 and 142.86 MHz. The latter is particularly important because it isolates a clock increase across the project latency boundary. Raising the clock is expected to reduce compute time approximately inversely with frequency when the accelerator is compute-bound, although DMA and software overhead do not scale in the same way.

\subsection{Measurement procedure}
Sustained throughput is measured with a batch of 100 images using the deployed FINN/PYNQ accelerator. Power is measured at the 12 V input rail and efficiency is computed as
\begin{equation}
\eta=\frac{\mathrm{FPS}}{P}\quad[\mathrm{FPS/W}].
\end{equation}
Both PL-only and end-to-end latencies are recorded. Detection accuracy is reported as mAP@0.5. Pascal VOC is used for the project closure criterion; filtered COCO results are shown as additional evidence where available. Because the supplied experiment table does not contain mAP for every hardware-only LP-YOLO clock/FIFO experiment, those rows are not treated as complete four-KPI candidates.

\section{NPU Implementations}
The experimental campaign is organized as a progression from a hardware baseline to a jointly optimized model.

\subsection{LP-YOLO 320 w4a4 with manual FIFO depth 6}
At 100 MHz, the baseline LP-YOLO implementation reaches 22.15 FPS at 2.00 W. Its PL latency is 62.41 ms and end-to-end latency is 66.47 ms. The design therefore demonstrates a functional streaming accelerator but misses the throughput and latency targets. Its efficiency, 11.075 FPS/W, is already above the 7 FPS/W requirement.

The same model and FIFO strategy are evaluated at 200 MHz. Throughput rises to 44.29 FPS and PL latency falls to 31.73 ms. Power increases only from 2.00 to 2.18 W, producing 20.32 FPS/W. This experiment demonstrates that clock frequency is a powerful lever for the lightweight architecture. It also illustrates why performance and accuracy experiments must be combined carefully: the supplied final table does not associate a VOC mAP with these two rows, so they cannot by themselves establish complete KPI closure.

\subsection{LP-YOLO 320 w4a4 with automatic FIFO sizing}
A second 100 MHz implementation replaces the manually fixed depth-6 FIFOs with FINN automatic FIFO sizing. Measured throughput rises to 35 FPS and PL latency decreases to 40.82 ms, while power is 2.15 W and efficiency is 16.28 FPS/W. Hardware performance therefore satisfies three hardware-oriented targets. However, the reported VOC mAP is 0.39 and COCO mAP is 0.27, both below the required VOC threshold. This implementation demonstrates that a fast accelerator is insufficient if the underlying detector does not preserve adequate accuracy.

\subsection{LP-YOLO Slim 256 w2a4 at 125 MHz}
The LP-YOLO Slim candidate combines architectural/input reduction with more aggressive mixed low-bit quantization. At 125 MHz it reaches 31.22 FPS at 2.80 W, corresponding to 11.15 FPS/W. VOC mAP is 0.594, so throughput, efficiency, and accuracy all pass. PL latency, however, is 51.49 ms, only 1.49 ms above the required limit. This result is important because it identifies a near-closure operating point without requiring a major architectural change.

\subsection{LP-YOLO Slim 256 w2a4 at 142.86 MHz}
The clock is increased from 125 to 142.86 MHz, while retaining the same model and quantization. Throughput increases to 35.66 FPS and PL latency falls to 45.11 ms. Power rises to 2.91 W, while efficiency improves to 12.25 FPS/W. Since the model is unchanged, VOC mAP remains 0.594. This is the decisive implementation because all four project objectives are simultaneously satisfied.

\subsection{YOLOv3-tiny custom w4a4}
The custom YOLOv3-tiny result provides the high-accuracy comparison. It reaches 0.761 VOC mAP and 0.559 COCO mAP, substantially above the lightweight variants. However, at 100 MHz it achieves only 12.35 FPS, consumes 2.08 W for 5.94 FPS/W, and has 94.87 ms PL latency. It therefore fails three of the four project requirements despite its superior detection accuracy.

\section{Experimental Results}
Table~\ref{tab:results} consolidates the measured results from the final comparison sheet. A dash denotes an accuracy value not supplied for that hardware experiment.

\begin{table*}[t]
\centering
\caption{Measured NPU implementations on PYNQ-Z1.}
\label{tab:results}
\resizebox{\textwidth}{!}{%
\begin{tabular}{lrrrrrrrrr}
\toprule
\textbf{Implementation} & \textbf{CLK} & \textbf{Batch} & \textbf{FPS} & \textbf{Power} & \textbf{FPS/W} & \textbf{E2E Lat.} & \textbf{HW-PL Lat.} & \textbf{mAP VOC} & \textbf{mAP COCO} \\
 & \textbf{(MHz)} & & & \textbf{(W)} & & \textbf{(ms)} & \textbf{(ms)} & & \\
\midrule
LP-YOLO 320 w4a4, FIFO depth 6 & 100 & 100 & 22.15 & 2.00 & 11.075 & 66.47 & 62.41 & -- & -- \\
LP-YOLO 320 w4a4, FIFO depth 6 & 200 & 100 & 44.29 & 2.18 & 20.317 & 35.48 & 31.73 & -- & -- \\
LP-YOLO 320 w4a4, FIFO automatic & 100 & 100 & 35.00 & 2.15 & 16.279 & 44.96 & 40.82 & 0.390 & 0.270 \\
\textbf{LP-YOLO Slim 256 w2a4} & \textbf{142.86} & \textbf{100} & \textbf{35.66} & \textbf{2.91} & \textbf{12.25} & \textbf{49.77} & \textbf{45.11} & \textbf{0.594} & \textbf{0.492} \\
YOLOv3-tiny custom w4a4 & 100 & 100 & 12.35 & 2.08 & 5.938 & 101.02 & 94.87 & 0.761 & 0.559 \\
\bottomrule
\end{tabular}}
\end{table*}

\subsection{Throughput and efficiency}
The throughput target separates the implementations into two groups. The 100 MHz manual-FIFO LP-YOLO and YOLOv3-tiny custom fail at 22.15 and 12.35 FPS, respectively. The 200 MHz LP-YOLO, automatic-FIFO LP-YOLO, and both Slim configurations exceed 30 FPS. The highest measured throughput is 44.29 FPS for LP-YOLO at 200 MHz.

Energy efficiency provides a different ranking. Five of the six implementations exceed 7 FPS/W. The 200 MHz LP-YOLO reaches 20.32 FPS/W, followed by automatic-FIFO LP-YOLO at 16.28 FPS/W. Both Slim implementations remain above 11 FPS/W despite their higher measured power. YOLOv3-tiny custom is the only row below the efficiency requirement, at 5.94 FPS/W. Thus, the high-capacity detector pays for its accuracy not only in latency and throughput but also in useful inferences per watt.

\subsection{Latency behavior}
The 50 ms PL-latency gate is met by LP-YOLO at 200 MHz (31.73 ms), automatic-FIFO LP-YOLO at 100 MHz (40.82 ms), and LP-YOLO Slim at 142.86 MHz (45.11 ms). The 125 MHz Slim design is particularly informative because it records 51.49 ms: the gap to the target is only 2.9\% of the 50 ms threshold. A 14\% clock increase to 142.86 MHz is sufficient to move the measured result comfortably below the gate.

End-to-end latency remains higher than PL latency because the PS, transfers, and software post-processing are still present. For the final 142.86 MHz design, the difference is 49.77-45.11=4.66 ms. At 125 MHz the corresponding difference is 56.13-51.49=4.64 ms. The near-constant value supports the interpretation that this component is largely outside the PL compute clock domain.

\subsection{Accuracy and the performance trade-off}
Accuracy exposes the central model-selection trade-off. Automatic-FIFO LP-YOLO provides strong hardware performance but only 0.39 VOC mAP. YOLOv3-tiny custom provides 0.761 VOC mAP but fails the hardware requirements. LP-YOLO Slim occupies the useful middle ground: its 0.594 VOC mAP clears the 0.55 gate while leaving enough computational headroom to exceed 30 FPS and approach the 50 ms latency boundary.

The COCO column should not be interpreted as the closure metric because the stated project objective is based on VOC. Nevertheless, it provides additional context: the Slim model reports 0.492 on the filtered COCO evaluation (board-test protocol, re-measured Aug.\ 2026), while YOLOv3-tiny custom reaches 0.559. The larger model therefore retains a modest accuracy advantage on this corpus when hardware constraints are ignored, though the gap is narrower than the VOC-versus-COCO gap for the Slim model itself, discussed in Section~\ref{sec:vocvscoco}.

\subsection{KPI closure}
Table~\ref{tab:kpi} shows the final candidate against the four explicit thresholds.
\begin{table}[h]
\centering
\caption{KPI closure for LP-YOLO Slim at 142.86 MHz.}
\label{tab:kpi}
\begin{tabular}{lccc}
\toprule
\textbf{KPI} & \textbf{Target} & \textbf{Measured} & \textbf{Status} \\
\midrule
Throughput & $>30$ FPS & 35.66 FPS & PASS \\
Efficiency & $>7$ FPS/W & 12.25 FPS/W & PASS \\
HW latency & $<50$ ms & 45.11 ms & PASS \\
mAP@0.5 VOC & $>0.55$ & 0.594 & PASS \\
\bottomrule
\end{tabular}
\end{table}

The margins relative to the thresholds are also useful. Throughput is approximately 19\% above the 30 FPS requirement. Energy efficiency is approximately 75\% above 7 FPS/W. PL latency is 4.89 ms below the maximum allowed value, approximately 10\% of the 50 ms threshold. VOC mAP exceeds the target by 0.044, or 8\% relative to the 0.55 threshold, making accuracy the tightest KPI, with latency a close second. The final solution therefore has substantial margin in efficiency and throughput, and a moderate but no longer marginal margin in latency, with accuracy remaining the tightest constraint.

\section{Discussion}
\subsection{Why the 142.86 MHz Slim design is the closing point}
The results show that no single optimization lever is sufficient. The 200 MHz LP-YOLO experiment proves that clock scaling can deliver excellent hardware performance, but the accuracy evidence for the lightweight 320 configuration does not establish the required 0.55 VOC mAP. The custom YOLOv3-tiny experiment proves that accuracy can be high, but its workload is too expensive to satisfy real-time and efficiency constraints. The Slim path succeeds because it first reduces the algorithmic workload and precision, then applies a hardware-frequency adjustment only after the model is already close to the target.

The 125-to-142.86 MHz comparison is therefore the most important local optimization in the final stage. The frequency increase is approximately 14\%, throughput rises from 31.22 to 35.66 FPS (14.2\%), and PL latency decreases from 51.49 to 45.11 ms (12.4\%). These changes are close to the inverse/linear scaling expected for a compute-dominated accelerator. Power increases by 3.9\% (2.80 to 2.91 W), so FPS/W improves from 11.15 to 12.25. This is a favorable scaling point because the extra power is converted into a disproportionate gain in throughput and latency margin while crossing the latency boundary.

\subsection{Role of FINN optimization}
FINN is central to the hardware exploration because it exposes parallelization and buffering as design parameters. PE and SIMD choices determine the folding of each layer and therefore the initiation rate of the pipeline. The project report notes that early, spatially wide layers receive higher parallelism while deeper layers can remain closer to PE/SIMD values of one. This is consistent with balancing the pipeline: the goal is not uniform replication but sufficient resources at the layers that otherwise dominate the cycle count.

FIFO sizing interacts with this balance. The measured automatic-FIFO 100 MHz result (35 FPS) is markedly faster than the manual depth-6 100 MHz result (22.15 FPS). Since other model-level variables are held in the same LP-YOLO w4a4 family, the comparison demonstrates that dataflow buffering can materially affect realized throughput. It also motivates treating FINN's FIFO optimization as part of design-space exploration rather than a fixed implementation detail.

\subsection{Quantization as a deployment enabler}
The resource analysis explains why low precision is necessary. Full YOLOv3-tiny needs 34.7 Mbit even with 4-bit weights, whereas LP-YOLO requires approximately 1.4 Mbit at 4 bits. For a FINN streaming architecture that aims to keep weights on chip, this difference changes the feasibility of the design. Quantization also lowers the arithmetic cost of the compute units, enabling more PE/SIMD parallelism within the LUT budget. The QAT evidence is therefore especially important: low precision would not be useful if it destroyed the detector's mAP.

The final w2a4 Slim result illustrates the combined effect. Two-bit internal weights are substantially more aggressive than the w4a4 LP-YOLO baseline, yet the reported VOC mAP is 0.594. This permits a small 256$\times$256 model to remain above the accuracy threshold while meeting the hardware constraints.

\subsection{Why VOC outperforms the filtered COCO corpus for this deployment}
\label{sec:vocvscoco}
The VOC-versus-COCO accuracy gap for the Slim model (0.594 versus 0.492 mAP@0.5, board-test protocol) is not attributable only to a general difficulty difference between the two evaluation sets. A direct measurement of box-side statistics from the project's own annotation files~\cite{vocdataset} shows that Pascal VOC vehicle boxes are 2--3$\times$ larger than the filtered COCO-derived corpus's: median box side 67.9~px at 320~px input (59.4~px at the deployed 256~px resolution) versus 21.1~px (16.9~px) for COCO. The gap narrows but persists when both corpora are restricted to the identical $\geq$12~px task-specification floor used throughout this work: 71.1~px versus 32.3~px median at 320~px.

This size gap is mechanistically relevant to the deployed architecture. For a heavily width- and bit-reduced detector (w2a4, width~0.35) at 256$\times$256 input, a 17--32~px COCO-derived object resolves to only a small number of feature-map cells at typical detector strides, whereas VOC's 54--71~px median sits comfortably mid-scale on multiple feature maps, offering a plausible mechanism for the measured board-test accuracy gap between the two corpora (0.594 VOC versus 0.492 COCO-filtered mAP@0.5) beyond a general harder-test-set effect.

Two further factors move in the same direction and are not fully separable from the size effect. COCO-derived images carry roughly twice as many vehicle instances per labelled image (mean 3.65 versus VOC's 1.82), increasing crowding and occlusion pressure on a low-capacity quantized detection head. VOC's own conversion pipeline also discards boxes flagged \emph{difficult} (heavily occluded or truncated) at source, a curation step with no equivalent applied to the COCO-derived corpus in this project, whose original occlusion/crowd annotations were not retained. The object-size advantage is real and substantial, but it should be read alongside these compounding factors rather than as the sole explanation for the corpus gap.

\subsection{Hardware latency versus system latency}
The project objective specifies PL latency below 50 ms, and the final design achieves 45.11 ms. End-to-end latency remains 49.77 ms. This distinction should be preserved in any interpretation of the result. The NPU itself satisfies the hardware-latency target with margin, but the complete application still contains approximately 4.66 ms of overhead outside the accelerator. Future optimization of DMA transfers, memory copies, preprocessing, decoding, and NMS could therefore reduce system latency further without changing the FINN datapath.

\subsection{Limitations}
Two limitations should be stated explicitly. First, mAP values are not supplied for every clock/FIFO-only hardware experiment, so those rows cannot be declared full four-KPI solutions. Second, VOC and filtered-COCO results are different evaluation corpora and should not be numerically merged; VOC is used here because it is the stated closure criterion.

\section{Conclusion}
This paper presented a complete NPU optimization path for real-time vehicle detection on the PYNQ-Z1. The methodology combines PyTorch training, Brevitas quantization-aware training, QONNX export, FINN streaming dataflow compilation, PE/SIMD folding, FIFO optimization, Vivado implementation, and physical PYNQ benchmarking. The design-space exploration demonstrates that algorithmic and hardware choices must be evaluated jointly: the most accurate YOLOv3-tiny custom configuration is too slow and inefficient, while faster LP-YOLO configurations do not provide the required accuracy in the supplied evaluation.

LP-YOLO Slim with 256$\times$256 input and w2a4 quantization provides the required compromise. At 125 MHz it already satisfies throughput, efficiency, and VOC accuracy, but its 51.49 ms PL latency misses the target. Increasing the PL clock to 142.86 MHz produces 35.66 FPS, 12.25 FPS/W, 45.11 ms PL latency, and 0.594 VOC mAP@0.5. Consequently, this implementation is the only evaluated configuration in the final comparison that satisfies all four stated objectives simultaneously.

The experimental sequence also provides reusable design guidance for small FPGA NPUs. First reduce the model and precision until an on-chip streaming implementation is feasible; next use FINN folding and FIFO sizing to balance the dataflow pipeline; then use moderate clock scaling to close the remaining performance gap; finally validate every KPI on the physical board rather than relying only on cycle estimates. In this project, that combined strategy turns a near-pass at 125 MHz into complete KPI closure at 142.86 MHz without sacrificing the required detection accuracy.

\section*{Acknowledgment}
This work has been funded by the Open Call 3 of dAIEdge, agreement number dAI3OC06[cite: 1]. dAIEdge is a project with grant agreement 101120726[cite: 1].

\end{document}